\documentclass[onecolumn,floatfix,superscriptaddress,a4paper,
               showkeys,nofootinbib,preprint]{revtex4}

\usepackage{graphicx}
\usepackage{amsmath}
\usepackage{hyperref}

\newcommand{\bs}{\boldsymbol}
\newcommand*{\dis}{\displaystyle}

\begin{document}

\title{Interacting Boson System at Finite Temperatures:
\newline
The treatment of the lattice QCD results}
\author{D. Anchishkin}
\affiliation{Bogolyubov Institute for Theoretical Physics, 03143 Kyiv, Ukraine}
\author{V. Gnatovskyy}
\affiliation{Bogolyubov Institute for Theoretical Physics, 03143 Kyiv, Ukraine}
\author{D. Zhuravel}
\affiliation{Bogolyubov Institute for Theoretical Physics, 03143 Kyiv, Ukraine}
 \author{V. Karpenko}
 \affiliation{Bogolyubov Institute for Theoretical Physics, 03143 Kyiv, Ukraine}



\begin{abstract}
We study an interacting relativistic system of charged bosons at
finite temperature and fixed isospin density within a
thermodynamically consistent mean-field approach.
The interaction is represented by repulsive $\varphi^4$ and $\varphi^6$
self-interactions, which generate a density-dependent effective mass of the
bosonic quasiparticles.
We formulate the thermodynamics in the Extended Canonical Ensemble,
obtained from the grand canonical description by a Legendre transformation
from the isospin chemical potential to the conserved isospin density.
In the condensed phase the chemical potential is constrained by
the Bose-condensation condition and is
therefore determined self-consistently through the effective mass.
We apply the formalism to a pion-like system at finite isospin and
clarify the relation between the isospin chemical-potential
conventions used for pion and quark degrees of freedom.
Using the convention adopted in lattice-QCD calculations, we compare the
pressure, isospin density, energy density, and trace anomaly with
lattice results at $T=120$ MeV.
With one constant fitted to the lattice pressure, each model
reproduces the lattice isospin density, energy density, and trace anomaly.
Both the $\varphi^4$ and $\varphi^6$
models reproduce the main features of the lattice data, with the
$\varphi^6$ interaction providing the better description.
The Extended Canonical Ensemble provides a convenient framework for
treating interacting particle-antiparticle systems across the transition
to the Bose-condensed phase.
\end{abstract}

\maketitle

\section{Introduction}
\label{sec1}

The thermodynamic properties of strongly interacting matter at finite
temperature and density are relevant for relativistic nucleus--nucleus
collisions and for the study of QCD matter under extreme conditions.
At sufficiently high temperature and density, QCD-motivated effective models
and lattice simulations predict substantial modifications of hadronic
properties associated with chiral symmetry restoration and deconfinement.
In the hadronic phase, pions constitute the lightest degrees of freedom and,
because they are bosons carrying isospin, their system can undergo Bose--Einstein
condensation at sufficiently large isospin density.
Recent years the subject of the asymmetric isospin systems attract a valuable
attention, see Refs.~\cite{son-2001,kogut-2001,toublan-2001}.

The thermodynamics of interacting bosons differs essentially from that of
an ideal gas.
In particular, the interaction modifies the quasiparticle mass and therefore
affects simultaneously the particle density, pressure, energy density, and the
condition for Bose condensation.
A thermodynamically consistent mean-field
treatment of interacting scalar bosons with repulsive self-interactions was
developed in Refs.~\cite{mishustin-anchishkin-2019,anch-gnat-kondakova-2025}.
In this approach the interaction generates a scalar-density dependent effective
mass and an excess-pressure contribution, whose relation guarantees
thermodynamic consistency.

In the present work we investigate an interacting relativistic
particle--antiparticle system at finite temperature and fixed isospin density.
We consider repulsive $\varphi^4$ and $\varphi^6$ interactions and formulate the
thermodynamics within the Extended Canonical Ensemble (ECE).
The ECE is obtained from the grand canonical description by a Legendre
transformation in which the conserved isospin density, rather than the isospin
chemical potential, is taken as the independent variable.
This formulation is convenient in the condensed phase, where the
chemical potential is no longer an independent variable but is constrained by
the condensation condition.

Finite isospin density is one of the few regimes of dense QCD free of the
fermion sign problem and therefore directly accessible to lattice Monte Carlo
simulations~\cite{son-2001}.
An additional issue arises when comparing the pion description with lattice-QCD
calculations.
Different normalizations of the isospin chemical potential are used in the
literature.
For the pion degrees of freedom, consistency of the isospin number and the
corresponding chemical potential gives
\begin{equation}
\mu_I^{(\pi)} \,=\, \mu_I^{(q)} = \mu_u - \mu_d \,.
\end{equation}
With this convention the condensation condition is $\mu_I=m_\pi$ for an ideal
pion gas.
Lattice calculations commonly employ instead the convention
(see Refs.~\cite{brandt-2016,brandt-2017,brandt-2018,brandt-2021,brandt-2022})
\begin{equation}
\mu_I^{\mathrm{lat}} \,=\, \frac{1}{2} (\mu_u - \mu_d) \,,
\end{equation}
for which the same physical condensation condition is expressed as
$\mu_I^{\mathrm{lat}} = m_\pi/2$.
We shall use the latter convention when comparing our results with the lattice
data \cite{brandt-2021,brandt-2022}.

The purpose of this work is therefore twofold.
First, we formulate a self-consistent description of the thermal and
Bose-condensed phases of an interacting relativistic boson system at fixed
isospin density.
Second, we apply the resulting equations of state to finite-isospin lattice-QCD
data and compare the performance of the $\varphi^4$ and $\varphi^6$ mean-field models.
In particular, we investigate the pressure, isospin density, energy density,
and trace anomaly and determine which interaction provides the more adequate
description of the lattice results.

A systematic low-energy description of QCD at finite isospin density is
provided by chiral perturbation theory, which predicts the onset of pion
condensation and the equation of state close to the threshold at low
temperatures~\cite{son-2001,kogut-2001,mammarella-2015,carignano-2017}.
The present approach is complementary to it: it is not a systematic
low-energy expansion, but a thermodynamically consistent mean-field model
with a single interaction constant, applied at finite temperature across
the Bose-condensation transition.

Another widely used description of hadronic matter is the hadron
resonance gas model~\cite{satarov-2009,vovchenko-2017,anch-vovchenko-2015},
which includes the full spectrum of hadrons and resonances.
Here we deliberately keep the description minimal and include only the
charged pion-like degrees of freedom built of $u$ and $d$ quarks.
Heavier states that couple to the isospin chemical potential, such as the
charged $\rho$ mesons and the kaons, which are present in the $2+1$-flavor
lattice simulations, are neglected: at the considered temperature their
thermal contributions are suppressed by their large masses and are expected
to be small compared with the pion contribution.

\section{Interacting scalar mesons: The mean-field approximation}
\label{sec:complex-scalar-lattice}

Let us start with the Lagrangian density for scalar-boson charged field $\hat\phi(x)$
\begin{equation}
\mathcal{L}(x) \,=\, \mathcal{L}_0 \,+\, \mathcal{L}_{\rm int} \,
=\, \partial_\mu \hat\phi^\dagger(x) \, \partial^\mu \hat\phi(x) - m^2\, \hat\phi^\dagger(x) \hat\phi(x) \,
+\, \mathcal{L}_{\rm int} \,,
\label{eq:lagrangian-lattice1}
\end{equation}
where $\mathcal{L}_0$ is a free part of the Lagrangian density.
The complex scalar field describes two charged states, a particle and an
antiparticle, which for the pion system are identified with $\pi^+$ and $\pi^-$;
the neutral pion, which does not couple to the isospin chemical potential,
is not included.
Here $x = (t, \bs r)$ is the time-space four-vector and
$\partial_\mu \partial^\mu  = \partial^2/\partial t^2 - \bs \nabla^2 $.

In the present work we consider the $\varphi^4$ and $\varphi^6$ models that realize a
selfinteracting repulsion between particles in the system, i.e. we assume the
following interaction Lagrangian densities, respectively,
\begin{equation}
\mathcal{L}_{\rm int}(\hat \sigma) \,
=\, - \frac{\lambda}{2} \, \left[\hat\phi^\dagger(x) \hat\phi(x)\right]^2 \,,
\qquad  {\rm or}  \qquad
\mathcal{L}_{\rm int}(\hat \sigma) \,
=\, - \frac{b}{3} \, \left[\hat\phi^\dagger(x) \hat\phi(x)\right]^3 \,,
\label{eq:lagrangian-lattice2}
\end{equation}
where the interaction constants are positive, $\lambda \ge 0$, $b \ge 0$.
The repulsion between pions is what stabilizes the condensate at a finite
isospin density~\cite{son-2001}; here this mechanism is implemented
in a thermodynamically consistent mean-field form.
Next, we consider the mean-field approximation expanding the interaction
Lagrangian around the mean value $\sigma$ defined as:
\begin{equation}
\sigma =\, \,\big\langle \hat\phi^\dagger(x) \hat\phi(x) \big\rangle \,,
\label{eq:sigma-lattice}
\end{equation}
where $\big\langle \ldots \big\rangle$ means a thermal averaging.
Indeed, one can use the interaction Lagrangian density in approximate way
taking just two first terms of the expansion of $\mathcal{L}_{\rm int}(\hat \sigma)$
around the mean value, see \cite{mishustin-anchishkin-2019}.
Following this procedure we define
\begin{eqnarray}
\mathcal{L}_{\rm int}(\hat \sigma) \, \approx \,
\mathcal{L}_{\rm int}^{^{\rm MF}}(\hat \sigma) \,
=\, \mathcal{L}_{\rm int}(\sigma) \,
+\, \frac{\partial \mathcal{L}_{\rm int}(\sigma)}{\partial \sigma} \,
(\hat \sigma - \sigma) \,,
\label{eq:int-lagrangian-mf-lattice1}
\end{eqnarray}
where $\hat \sigma = \hat\phi^\dagger(x) \hat\phi(x)$ is the operator of the scalar density.

\subsection{The mean-field approximation}
\label{sec:mf-approximation}
Let us consider the mean-field (MF) Lagrangian density
\begin{equation}
\mathcal{L} \approx
\mathcal{L}_{_{\rm MF}} = \mathcal{L}_0 + \mathcal{L}_{\rm int}^{^{\rm MF}} \,,
\label{eq:lagrangian-mf-5a}
\end{equation}
where $\mathcal{L}_0$ is defined in eq.~(\ref{eq:lagrangian-lattice1})
and $\mathcal{L}_{\rm int}^{^{\rm MF}}$ is defined in
eq.~(\ref{eq:int-lagrangian-mf-lattice1}).
In accordance with these definitions we get
\begin{equation}
\mathcal{L}_{_{\rm MF}} \,=\, \partial_\mu \hat\phi^\dagger(x) \, \partial^\mu \hat\phi(x) \,
-\, M^2(\sigma)\, \hat\phi^\dagger(x) \hat\phi(x) \,+\, P_{\rm ex}(\sigma) \,,
\label{eq:lagrangian-mf-5}
\end{equation}
where
\begin{equation}
M^2(\sigma) =   m^2 + U(\sigma) \,,
\quad
U(\sigma) = - \frac{\partial \mathcal{L}_{\rm int}(\sigma)}{\partial \sigma} \,,
\quad
P_{\rm ex}(\sigma) = \mathcal{L}_{\rm int}(\sigma)
- \sigma \frac{\partial \, \mathcal{L}_{\rm int}(\sigma)}{\partial \sigma} \,.
\label{eq:definitions}
\end{equation}
and
\begin{equation}
P_{\rm ex}(\sigma) = \mathcal{L}_{\rm int}(\sigma)
- \sigma \frac{\partial \, \mathcal{L}_{\rm int}(\sigma)}{\partial \sigma} \,.
\label{eq:definition-pex}
\end{equation}
As it is seen from eq.~(\ref{eq:lagrangian-mf-5}), in the MF
approximation the scalar field ``earns'' an effective mass $M(\sigma)$
and, as it will be seen further, the quantity $P_{\rm ex}$ plays a role of
an excess pressure.
It is easy to check that in accordance with definitions (\ref{eq:definitions})
and (\ref{eq:definition-pex}) we get the following relation:
$\sigma (\partial U/\partial \sigma) = \partial P_{\rm ex}/\partial \sigma$,
which is a necessary attribute to provide the thermodynamic consistency of
the model.
For the two interactions considered here, $U(\sigma) = \lambda \sigma$
for the $\varphi^4$ model and $U(\sigma) = b \sigma^2$ for the $\varphi^6$ model.

\subsection{The condensed phase}
\label{sec:mf-condensate-phase}

In what follows we use the Bogolyubov's decomposition of the boson field
operator \cite{bogolyubov}
\begin{equation}
\hat \phi \,=\, \Phi_0 \,+\, \hat \psi
\qquad   {\rm with}  \qquad
\big\langle \hat \psi \big\rangle \,=\, 0 \,,
\label{eq:repres-phi}
\end{equation}
where $\Phi_0$ is the wave function which is responsible for the condensate
creation in the system and $\hat\psi$ is the field operator that describes the
thermal (kinetic) particles.

The field operators $\hat \phi(x)$ and
$\hat \pi(x) = \partial \mathcal{L}/\partial (\partial_0 \hat \phi)
= \partial \hat \phi^\dagger(x)/\partial x^0$
satisfy the following equal-time commutation relations
\begin{equation}
\left[\hat \phi(t,\bs r),\, \hat \pi(t,\bs r')\right]\,
=\, i \delta^3(\bs r - \bs r')
\quad \Rightarrow \quad
\left[\hat \psi(t,\bs r),\, \frac{\partial \hat \psi^\dagger(t,\bs r')}{\partial t}\right]
=\, i \delta^3(\bs r - \bs r')\,.
\label{eq:quantization}
\end{equation}
Hence, the thermal field $\hat \psi(x)$ satisfies the same equal-time
commutation relations as the total scalar field $\hat \phi(x)$.
Using this we can write the Bogolyubov's decomposition in the relativistic case
as
\begin{eqnarray}
\hat \phi(t,{\bf r}) \,=\,
\Phi_0(t) \,+\,
 \int_{\bs p \neq 0} \frac{d^3p}{(2\pi)^3 2\omega_p} \left[ a_{\bs p}e^{-i \omega_p t}
 + b^\dagger_{\bs p}e^{i \omega_p t}\right] e^{i{\bf p}\cdot {\bf r}} \,,
 \label{eq:expansion-phi-ideal-1}
 \end{eqnarray}
where $\omega_p = \sqrt{M^2 + {\bs p}^2}$ and we assume that the condensate
medium is spatially homogeneous.
Then, in accordance with equal-time commutation relations (\ref{eq:quantization}),
the boson annihilation and creation operators,
$a_{\bs k}$, $a^\dagger_{\bs k}$ and $b_{\bs k}$, $b^\dagger_{\bs k}$,
satisfy the commutation relations
\begin{equation}
\big[ \ a_{\bs k},\, a^\dagger_{\bs k'} \big] \,=\, (2\pi)^3\,2\omega_k \,
\delta^3(\bs k - \bs k') \,,
\qquad
\big[ b_{\bs k},\, b^\dagger_{\bs k'} \big] \,=\, (2\pi)^3\,2\omega_k \,
\delta^3(\bs k - \bs k') \,.
\label{eq:com-rel-ideal}
\end{equation}
In what follows we investigate the bosonic systems in thermodynamic
limit $V \to \infty$, that is instructive to fix the following
relations
\begin{equation}
\int d^3r \, \hat\psi(t, \bs r) \,=\, 0 \,,  \quad  \quad
\int d^3r \, \partial^\mu \hat\psi(t, \bs r) \,=\, 0 \,.
\label{eq:zero-int-psi}
\end{equation}
With account to these relations we integrate the MF Lagrangian density
(\ref{eq:lagrangian-mf-5}) over the spatial volume occupied by the system,
$L_{_{\rm MF}}(t) = \int d^3r \mathcal{L}_{_{\rm MF}}(t,\bs r)$, and obtain
\begin{eqnarray}
L_{_{\rm MF}}(t)
=  V \left[\partial_t \Phi_0^* \partial_t \Phi_0 - M^2 \Phi_0^*\Phi_0\right]
+ \! \int \! d^3r \left[\partial_\mu \hat\psi^\dagger(x) \partial^\mu \hat\psi(x)
- M^2\hat\psi^\dagger(x) \hat\psi(x)\right] + V P_{\rm ex}(\sigma) .
\label{eq:lagrangian-ideal-3}
\end{eqnarray}
All terms in (\ref{eq:lagrangian-mf-5}) that are linear on the thermal scalar
field $\hat\psi$ disappear after integration over the spatial volume in
accordance with eqs.~(\ref{eq:zero-int-psi}).
However, it will be not true in case if the scalar condensate field $\Phi_0$
depends on spatial coordinates (these statements can be also found in
Ref.~\cite{kiselev-2000}).

The action is a functional of two fields
$S[\Phi_0,\hat\psi] =  \int dt L_{_{\rm MF}}(t;\sigma)$.
Using the MF Lagrangian (\ref{eq:lagrangian-ideal-3}) one can get
equations of motion for the condensate field $\Phi_0$ and thermal field $\hat\psi$
by minimizing the action
\begin{eqnarray}
\label{eq:eq-phi0-20}
\left[ \partial_0 \partial^0 + M^2(\sigma) \right] \Phi_0 &=& 0 \,,
\\
\left[\partial_\mu \partial^\mu + M^2(\sigma)\right] \hat\psi(x) &=& 0 \,.
\label{eq:eq-psi-20}
\end{eqnarray}
There are two solutions of eq.~(\ref{eq:eq-phi0-20}).
The first one is trivial $\Phi_0 = 0$, it corresponds to an absence of
the condensate in the system.
The second solution indicates the presence of the condensate
\begin{eqnarray}
\Phi_0 \,=\, \frac{e^{-iMt}}{\sqrt{2M}} \, \psi_0^{(+)} +
\frac{e^{iMt}}{\sqrt{2M}} \, \left(\psi_0^{(-)}\right)^* \,.
\label{eq:eq-phi0-40}
\end{eqnarray}
We would like to stress that $\psi_0^{(+)}$ and $\psi_0^{(-)}$ are the
nonrelativistic wave functions attributed to particles and antiparticles,
which are in rest.

In accordance with the MF Lagrangian (\ref{eq:lagrangian-ideal-3})
the effective Hamiltonian in the MF approach
$H_{_{\rm MF}}(\sigma) = \int d^3r \mathcal{H}_{_{\rm MF}}(x,\,\sigma)$
has a quasi-free form
\begin{equation}
H_{_{\rm MF}} = V \left[\partial_t \Phi_0^* \partial_t \Phi_0
+ M^2(\sigma) \Phi_0^*\Phi_0\right]
+ \int d^3r \! \left[\partial_t \hat\psi^\dagger \partial_t \hat\psi
+ \bs \nabla \hat\psi^\dagger \cdot \bs \nabla \hat\psi
+ M^2(\sigma) \hat\psi^\dagger \hat\psi \right] - V P_{\rm ex}(\sigma) .
\label{eq:hamiltonian-11}
\end{equation}
Applying solution (\ref{eq:eq-phi0-40}), we obtain the condensate contribution
to the MF Hamiltonian:
\begin{equation}
\partial_0 \Phi_0^* \, \partial_0 \Phi_0 + M^2 \Phi_0^* \Phi_0 \,
=\, M\, \left[\left(\psi_0^{(+)}\right)^*\psi_0^{(+)} \,
+\, \left(\psi_0^{(-)}\right)^*\psi_0^{(-)}\right] \,=\, M n_{\rm cond} \,,
\label{eq:def-psi0-10}
\end{equation}
where the condensate particle-number density $n_{\rm cond}$ reads:
\begin{equation}
n_{\rm cond} \,=\, \left(\psi_0^{(+)}\right)^*\psi_0^{(+)} \, +\,
\left(\psi_0^{(-)}\right)^*\psi_0^{(-)} \,.
\label{eq:def-psi0-20}
\end{equation}
In correspondence to eq.~(\ref{eq:repres-phi}) the total scalar density
$\sigma = \langle \hat \phi^\dagger \hat \phi\rangle$ also can be decomposed as:
\begin{equation}
\sigma = \sigma_{\rm cond} + \sigma_{\rm th},
\qquad {\rm where} \qquad
\sigma_{\rm cond} = \Phi_0^* \Phi_0 \,, \quad
\sigma_{\rm th} = \big\langle \hat\psi^\dagger(x) \hat\psi(x) \big\rangle \,.
\label{eq:averagings-2}
\end{equation}
The connection between the condensate scalar density and the condensate
particle-number density is given by
\begin{equation}
\sigma_{\rm cond} \,
=\, \frac{1}{2M}\, \left[\left(\psi_0^{(+)}\right)^*\psi_0^{(+)} \,
+\, \left(\psi_0^{(-)}\right)^*\psi_0^{(-)}\right] \,
=\, \frac{\dis n_{\rm cond}}{\dis 2M} \,.
\label{eq:relation}
\end{equation}
Then, one can represent the MF Hamiltonian
(\ref{eq:hamiltonian-11}) in terms of creation and annihilation operators as:
\begin{equation}
H_{_{\rm MF}}(\sigma) \,=\, M V n_{\rm cond} + \frac 12 \int \frac{d^3k}{(2\pi)^3} \,
\big[ a^\dagger(\bs k) a(\bs k)\, +\, b^\dagger(\bs k) b(\bs k) \big] \,
-\, V\,P_{\rm ex}(\sigma) \,.
\label{eq:tot-hamiltonian2}
\end{equation}
Averaging of the MF Hamiltonian (\ref{eq:hamiltonian-11}) gives the energy
density of the system, $\varepsilon = \langle H \rangle/V$:
\begin{eqnarray}
\varepsilon \,=\, M n_{\rm cond}
+ \frac 1V \int d^3r \! \left\langle \partial_t \hat\psi^\dagger \partial_t \hat\psi
+ \bs \nabla \hat\psi^\dagger \cdot \bs \nabla \hat\psi \right\rangle
+ m^2 \sigma_{\rm th} +  U(\sigma)\sigma_{\rm th} -  P_{\rm ex}(\sigma) \,.
\label{eq:hamiltonian-11a}
\end{eqnarray}

\subsection{Isospin density in bosonic system }
\label{sec:isospin-density-operator}

The Noether current, which guarantees the isospin conservation, in the operator
form reads $\hat j^{\mu}(x) = i \left[\hat\phi^\dagger(x) \partial^\mu \hat\phi(x)
- \left(\partial^\mu \hat\phi^\dagger(x) \right) \hat\phi(x)   \right]$.
The operator $\hat N_I$ of the total isospin number of the system is obtained
as an integral of the zero component $\hat j^0$ of the Noether current over the
volume of the system:
\begin{equation}
\hat N_I(t,\bs r)\,=\, i \int d^3r \,
\left[\hat\phi^\dagger(t,\bs r) \partial_t \hat\phi(t,\bs r) \,
-\, \left(\partial_t \hat\phi^\dagger(t,\bs r) \right)\, \hat\phi(t,\bs r) \right] \,.
\label{eq:integral-zero-current}
\end{equation}
Using decomposition $\hat \phi = \Phi_0 + \hat \psi$
one can
show that in a homogeneous system the isospin-number operator $\hat N_I$ consists
of the condensate $N_{I\rm cond}$ and thermal $\hat N_{I\rm th}$ contributions
\begin{equation}
\hat N_I \,=\, N_{I\rm cond} +  \hat N_{I\rm th} \,,
\label{eq:isospin-operator-1}
\end{equation}
where using solution (\ref{eq:eq-phi0-40}) we get
\begin{eqnarray}
N_{I\rm cond} \,=\, V\,\left[\left(\psi_0^{(+)}\right)^*\psi_0^{(+)} \,
-\, \left(\psi_0^{(-)}\right)^*\psi_0^{(-)}\right] \,=\, V n_{I\rm cond}\,.
\label{eq:isospin-cond-2}
\end{eqnarray}
(It is instructive to compare the expression obtained for $n_{I\rm cond}$ with
$n_{\rm cond}$ given in eq.~(\ref{eq:def-psi0-20}).)
The thermal contribution $\hat N_{I\rm th}$ to the isospin density  in
eq.~(\ref{eq:isospin-operator-1}) is given by
\begin{equation}
\hat N_{I\rm th} = \hat N^{(+)}_{\rm th} - \hat N^{(-)}_{\rm th} \,,
\quad {\rm where} \quad
\hat N^{(+)}_{\rm th} \,=\, \int \frac{d^3k}{(2\pi)^3 2\omega_k}\,
a^\dagger_{\bs k}\, a_{\bs k}\,,
\quad
\hat N^{(-)}_{\rm th} \,=\, \int \frac{d^3k}{(2\pi)^3 2\omega_k}\,
b^\dagger_{\bs k}\, b_{\bs k}\,.
\label{eq:oper-particle-dens}
\end{equation}

\section{Bosonic system of particles and antiparticles within the Extended
Canonical Ensemble}
\label{sec:thermal-cond-phase}

We are going to describe thermodynamic properties of the boson system of
particles and antiparticles in the framework of the Extended Canonical
Ensemble (ECE) that is obtained after movement from the Grand Canonical
Ensemble (GCE) to Canonical Ensemble (CE) by means of the Legendre transformation,
see Ref.~\cite{anch-gnat-kondakova-2025}.
Our starting point is the partition function in the GCE, which
in the thermal phase and in the condensed phase is given by, respectively,
\begin{eqnarray}
\label{eq:gce-partition-func-therm}
Z(T,\mu_I,V) &=&
e^{ \beta V\,P_{\rm ex}(\sigma)} \,
{\rm Tr}\left\{ e^{- \beta \left[H_{_{\rm MF}} - \mu_I \hat N_I \right]_{\rm th}}\right\} \,,
\\
Z(T,\mu_I,V) &=&
e^{- \beta \left[N_{\rm cond}(M - \mu_I) - V\,P_{\rm ex}(\sigma)\right]} \,
{\rm Tr}\left\{ e^{- \beta \left[H_{_{\rm MF}} - \mu_I \hat N_I \right]_{\rm th}}\right\} \,,
\label{eq:charge-id-gce-part-func-cond-2}
\end{eqnarray}
where we introduce notation for thermal quantities:
\begin{equation}
\left[H_{_{\rm MF}} - \mu_I \hat N_I \right]_{\rm th}
=  \int \frac{d^3k}{(2\pi)^3 2\omega_k} \,
\left[(\omega_k - \mu_I) a^\dagger_{\bs k} a_{\bs k}
+ (\omega_k + \mu_I) b^\dagger_{\bs k} b_{\bs k}\right] \,.
\label{eq:charge-id-gce-part-func-cond-3}
\end{equation}
Note, the C-number contributions to the Hamiltonian (\ref{eq:tot-hamiltonian2})
and particle-number operator (\ref{eq:isospin-operator-1}) associated with
the presence of the condensate
do not affect the mean value of some operator $\hat A$ in comparison with the
statistical averaging of this operator in the thermal phase.

After calculation of the partition function one can write the
thermodynamic potential $\Omega = - T \, \ln{ Z(T,\mu_I;\sigma) }$
in a homogeneous thermal system and in condensate system, respectively,
\begin{eqnarray}
\label{eq:omega-therm}
& \Omega(T,\mu_I,V) = \Omega_{\rm th}(T,\mu_I,V) \,-\, VP_{\rm ex}(\sigma) \,,
\quad &{\rm thermal \ phase}
\\
& \Omega(T,\mu_I,V) =
N_{\rm cond}(M - \mu_I) \,+\, \Omega_{\rm th}(T,\mu_I,V) - VP_{\rm ex}(\sigma) \,,
\quad &{\rm condensed \ phase}
\label{eq:omega-cond-1}
\end{eqnarray}
where
\begin{eqnarray}
\label{eq:omega-cond-2}
\Omega_{\rm th}(T,\mu_I,V) \,=\, VT \int \frac{d^3k}{(2\pi)^3}
\left\{\ln{ \left[ 1- e^{- \big(\omega_k(\sigma) - \mu_I \big)/T } \right] }
+ \ln{ \left[ 1 - e^{- \big(\omega_k(\sigma) + \mu_I \big)/T } \right] }\right\} \,.
\end{eqnarray}
The condensate term in eq.~(\ref{eq:omega-cond-1}) vanishes identically once
the onset condition $\mu_I = M$ is imposed, but it has to be retained in the
variational procedure that determines $n_{\rm cond}$.

In the GCE, the isospin number $n_I = N_I/V$ is related to the corresponding
chemical potential $\mu_I$ as:
\begin{equation}
n_I \,=\, - \, \frac 1V \frac{\partial \Omega}{\partial \mu_I} \,.
\label{eq:charge-legendre-transform-1}
\end{equation}
We use this relation as a basis for the Legendre transform:
after solving this equation with respect to the chemical potential one obtains
dependence $\mu_I = \mu_I(T,n_I)$, i.e.,
instead of the variable $\mu_I$ we go to the new variable $n_I$.
It is analog of the definition of the ``effective potential'' that is the
standard transformation in the quantum field theory.
In our case the ``effective potential'' will be the extended free energy.
Making the Legendre transformation we move from the Grand Canonical Ensemble
to Extended Canonical Ensemble, where we define the free energy as
\begin{equation}
F(T,N_I,V) \,=\, \Omega \,-\, \mu_I \, \frac{\partial \Omega}{\partial \mu_I}\,.
\label{eq:charge-cond-legendre-transform-2}
\end{equation}
Here we imply that the chemical potential is determined by $T$ and $n_I$.
From now on, the statistical operator $\hat \rho$ and the corresponding partition
function in the ECE are defined as
\begin{equation}
\hat \rho = e^{- \beta \left[H_{_{\rm MF}} - \mu_I(T,n_I) \hat N_I \right] } \,,
\qquad
Z(T,n_I) \,=\, {\rm Tr} \, {\hat \rho} \,.
\label{eq:ece-partition-func-2}
\end{equation}
And in place of the connection between the partition function and the free energy
in the Canonical Ensemble, $F = -T\ln{ {\rm Tr}\, e^{-\beta H}}$, in the Extended
Canonical Ensemble in accordance with transformation
(\ref{eq:charge-cond-legendre-transform-2}) the free energy is given by
\begin{equation}
F(T,N_I,V) \,=\, N_I \mu_I(T,n_I) \,-\, T \ln{{\rm Tr}\left\{ e^{- \beta \left[H
- \mu_I(T,n_I) \hat N_I \right] } \right\} } \,,
\label{eq:ece-free-energy-1}
\end{equation}
where $\mu_I = \mu_I(T,n_I)$ is solution of eq.~(\ref{eq:charge-legendre-transform-1}).

Using eq.~(\ref{eq:omega-cond-1}) for the condensed phases
we can rewrite eq.~(\ref{eq:charge-legendre-transform-1}) in explicit form
\begin{eqnarray}
\label{eq:id-charge-density}
n_I &=& n_{\rm cond} + n_{I \rm th} \,,
\\
n_{I\rm th}(T,\mu_I) &=& \int \frac{d^3k}{(2\pi)^3}\,
\left[f_{\rm BE}\big(\omega_k(\sigma), \mu_I\big)
- f_{\rm BE}\big(\omega_k(\sigma),- \mu_I\big) \right] \,,
\label{eq:charge-density-th}
\end{eqnarray}
where $\omega_k(\sigma) = \sqrt{M(\sigma)^2+\bs k^2}$, $M^2 = m^2 + U(\sigma)$
and the Bose-Einstein distribution function is given by
$f_{_{\rm BE}}(\omega,\mu) = \{ \exp{[(\omega - \mu)/T]} - 1 \}^{-1}$.

At this point we have to make a very important note.
Let us assume that both components of the Bose gas are in the condensed phase.
Then, in accordance with the condition for condensate formation, for the arguments
of the distribution functions in eq.~(\ref{eq:id-charge-density}) should be valid
two equalities
\begin{eqnarray}
M(\sigma) \,-\, \mu_I \,=\, 0 \,,
\qquad
M(\sigma) \,+\, \mu_I \,=\, 0 \,.
\label{eq:cond-condensate}
\end{eqnarray}
Evidently this set of equations leads to $M(\sigma) = 0$ and $\mu_I = 0$,
simultaneously.
The first equality, i.e. a zero effective mass of the quasiparticles, cannot
be realized for the system with only a repulsion between particles,
$M^2 = m^2 + U(\sigma)$, where $U(\sigma) \ge 0$, see eq.~(\ref{eq:definitions}).
Therefore, only one component of the system can develop a condensate when we
keep $n_I \ne 0$.\footnote{If a system, in addition to repulsive
interactions, possesses a strong attractive interaction, it becomes possible.}
Let it be the particles, i.e. the component that carries the positive
isospin charge and enters with $+\mu_I$ (for the pion system these are the
$\pi^+$ mesons), for which the first equation in (\ref{eq:cond-condensate})
is valid, $\mu_I = M(\sigma)$ in the condensed phase.
The latter means that the antiparticles ($\pi^-$ mesons) exist only in the
thermal phase for all temperature range, hence, they cannot be found in the
condensed phase, i.e., the antiparticle amplitude vanishes,
$\psi_0^{(-)} = 0$ in eq.~(\ref{eq:eq-phi0-40}).

So, making the choice of the positive isospin charge of the dominant component
in the pion gas we obtain equations $n_I = n_{I\rm th}(T,\mu_I)$ for the thermal
phase (see eq.~(\ref{eq:id-charge-density})), and
$n_I = n_{\rm cond} + n_{I\rm th}(T,M)$, where $\mu_I = M(\sigma)$, for the
condensed phase.
We use these equations separately in both phases to determine the chemical
potential as a function of the temperature and isospin number.
Therefore, in the thermal phase the chemical potential as the solution of
equation $n_I = n_{I \rm th}(T,\mu_I)$
is a function of the temperature and the isospin density, $\mu_I = \mu_I(T,n_I)$.
However, in the condensed phase the value of the chemical potential,
$\mu_I = M(\sigma)$, depends on the temperature and the isospin density via a
dependence on the scalar density, $\sigma = \sigma(T,n_I)$.

Using eqs.~(\ref{eq:omega-therm}) and (\ref{eq:omega-cond-1}) one can write
the Legendre transformation (\ref{eq:charge-cond-legendre-transform-2})
in explicit form for the thermal and condensed phases, respectively:
\begin{eqnarray}
\label{eq:free-energy-1}
\mathcal{F}(T,n_I) &=&  \mu_I \, n_I + \mathcal{F}_{\rm th}(T,n_I) - P_{\rm ex} \,,
\\
\mathcal{F}(T,n_I) &=& M \, n_I + \mathcal{F}_{\rm th}(T,n_I) - P_{\rm ex} \,,
\label{eq:cond-free-energy-1}
\end{eqnarray}
where the thermal (kinetic) contribution is
\begin{eqnarray}
\mathcal{F}_{\rm th}(T,n_I) =  T \int \frac{d^3k}{(2\pi)^3} \left\{
\ln{ \left[ 1- e^{- \big(\omega_k - \mu_I \big)/T } \right] }
+ \ln{ \left[ 1- e^{- \big(\omega_k + \mu_I \big)/T } \right] } \right\} \,,
\label{eq:free-energy-th}
\end{eqnarray}
where $\mathcal{F}(T,n_I) = F(T,N_I,V)/V$ is the density of the free energy,
and we recall that in the thermal phase $\mu_I = \mu_I(T,n_I)$ and in the
condensed phase $\mu_I = M\large(\sigma(T,n_I)\large)$.

\subsection{Self-consistent equations}
\label{sec:self-consistent-equations}

For a closed description of the system it is necessary
to formulate the self-consistent equations that determine a dependence of the
thermodynamic quantities on the free variables $T$ and $n_I$.
In thermodynamically equilibrium case we use the Bose-Einstein distribution
functions that were determined above.
In the thermal phase we return to equation
$n_I = n_{I \rm th}(T,\mu_I)$, see (\ref{eq:id-charge-density}).
We calculate also the thermal scalar density, see eq.~(\ref{eq:averagings-2}),
as a derivative of the $\Omega_{\rm th}$, see eq.~(\ref{eq:omega-cond-2}),
with respect to $M^2$:
$\sigma = \langle \hat \psi^\dagger \hat \psi \rangle
= (1/V) \partial \Omega_{\rm th}/\partial M^2$.
Or calculate it directly using the averagings:
$\langle a^\dagger_{\bs k} a_{\bs k} \rangle = 2V \omega_k f_{_{\rm BE}}(\omega_k,\mu_I)$
and
$\langle b^\dagger_{\bs k} b_{\bs k} \rangle = 2V \omega_k f_{_{\rm BE}}(\omega_k,-\mu_I)$.
As a result we formulate self-consistent set of equations
\begin{eqnarray}
&& n_I  \,=\,   \int \frac{d^3k}{(2\pi)^3}\,
\left[ f_{\rm BE}\big(\omega_k,\mu_I\big) - f_{\rm BE}\big(\omega_k,-\mu_I\big)
\right] \,,
\label{eq:isospin-density}
\\
&& \sigma \,=\,  \int \frac{d^3k}{(2\pi)^3\, 2\omega_k}\,
\left[ f_{\rm BE}\big(\omega_k,\mu_I\big) + f_{\rm BE}\big(\omega_k,-\mu_I\big)
\right] \,,
\label{eq:total-scalar-density}
\end{eqnarray}
where $\sigma$ is the total thermal scalar density,
$\omega_k = \sqrt{M(\sigma)^2 + \bs k^2}$ and $M^2(\sigma)  =  m^2 + U(\sigma)$.
This set of equations must be solved for a fixed temperature $T \ge T_{\rm c}$
and isospin density $n_I$.
Finally, we get the functions $\mu_I(T,n_I)$ and $\sigma(T,n_I)$.

In the condensed phase the thermodynamic state of
the system is determined by a self-consistent equation for the scalar density,
$\sigma =  \sigma_{\rm cond} + \sigma_{\rm th}$, which
looks like eq.~(\ref{eq:total-scalar-density}) that is completed by the
contribution of the condensate scalar density.
And we also remember, that according to eq.~(\ref{eq:relation}) there is relation
$\sigma_{\rm cond} = n_{\rm cond}/2M$ (recall that antiparticles, $\pi^-$ mesons,
are only in the thermal phase).
Therefore, in the condensed phase the thermodynamic state of the system is
determined by a set of two equations for $\sigma$ and $n_{\rm cond}$:
\begin{eqnarray}
\label{eq:charge-density-3}
&& n_I = n_{\rm cond}(T,n_I)
+ \int \frac{d^3k}{(2\pi)^3}\, \left[f_{\rm BE}\big(\omega_k, M(\sigma)\big)
- f_{\rm BE}\big(\omega_k(\sigma),- M(\sigma)\big) \right] \,,
\\
&& \sigma =  \frac{\dis n_{\rm cond}}{\dis 2M}
+ \int \frac{d^3k}{(2\pi)^3\, 2\omega_k}\, \left[f_{\rm BE}\big(\omega_k, M(\sigma)\big)
+ f_{\rm BE}\big(\omega_k,- M(\sigma) \big) \right] \,,
\label{eq:total-scalar-density-cond-2}
\end{eqnarray}
where, as in the thermal phase,
$\omega_k = \sqrt{M^2(\sigma) + \bs k^2}$ and $M^2(\sigma)  =  m^2 + U(\sigma)$.
This set of equations must be solved for a fixed temperature, $T$, and isospin
density $n_I$, then we obtain functions:
$\sigma(T,n_I)$, $\mu_I(T,n_I) = M(\sigma(T,n_I))$, $n_{\rm cond}(T,n_I)$ and
$\sigma_{\rm cond}(T,n_I)$.

The pressure and energy density are related to the free energy density as:
\begin{eqnarray}
\label{eq:f-1a}
p = n_I\, \mu_I(T,n_I) \,-\, \mathcal{F}(T,n_I)\,,
\qquad
\varepsilon = \mathcal{F}(T,n_I) \,
-\, T\, \left[\frac{\partial \mathcal{F}(T,n_I)}{\partial T}\right]_{n_I} \,.
\label{eq:f-2a}
\end{eqnarray}
Using eqs.~(\ref{eq:free-energy-1}) and (\ref{eq:cond-free-energy-1})
we get the pressure
\begin{eqnarray}
\label{eq:pressure-cond}
p(T,n_I) =  - \mathcal{F}_{\rm th}(T,n_I) + P_{\rm ex}(\sigma) \,,
\label{eq:pres-2}
\end{eqnarray}
where $\mathcal{F}_{\rm th}(T,n_I)$ is given in eq.~(\ref{eq:free-energy-th}).
The integral on the r.h.s. of this equation gives the pressure of thermal
quasi particles and antiparticles, what together
can be treated as the kinetic pressure of the quasi-ideal gas.
The additional pressure $P_{\rm ex}$ is due to interaction in the system.
From this decomposition we see that the quantity $P_{\rm ex}$ is indeed the
excess pressure.
The condensate itself does not produce an independent kinetic-pressure
contribution.

For calculation of the energy density we use eq.~(\ref{eq:f-2a}) and obtain
for the thermal and condensed phases, respectively:
\begin{eqnarray}
\label{eq:tot-energy-2}
\varepsilon(T,n_I) &=&  \int \frac{d^3k}{(2\pi)^3} \, \omega_k \,
\left[ f_{\rm BE}(\omega_k, \mu_I)  \,
+\, f_{\rm BE}(\omega_k,- \mu_I)\right] \,-\, P_{\rm ex}(\sigma) \,.
 \\
\varepsilon(T,n_I) &=& M n_{\rm cond} \,+\, \int \frac{d^3k}{(2\pi)^3} \, \omega_k \,
\left[ f_{\rm BE}(\omega_k,M)  \,+\, f_{\rm BE}(\omega_k,- M)\right] \,
-\, P_{\rm ex}(\sigma) \,,
\label{eq:tot-energy-cond-2}
\end{eqnarray}
where $\mu_I = \mu_I(T,n_I)$ in the thermal phase.

\section{The quark content of the pionic phase}
\label{sec:quark-content}

Our aim in this section is to put in correspondence the definitions of the
isospin chemical potentials $\mu_I^{(\pi)}$ and $\mu_I^{(q)}$ for pion
and for quark degrees of freedom, respectively.

Let us assume that the hadron system is represented by pions as the most light
entities, and the baryon charge of the system is zero, $N_{{\rm B}} = 0$.
Then, for the pion hadron phase the isospin number reads
(recall $I_\pi = 1$, with three projections $0, \pm 1$)
\begin{equation}
N_I \,=\, N_\pi^{(+)} - N_\pi^{(-)} \,,
\label{eq:ni-pions-1}
\end{equation}
where $N_\pi^{(+)}$ and $N_\pi^{(-)}$ are the particle numbers of $\pi^+$ and
$\pi^-$ mesons, respectively.
In what follows we identify the condensing component with the positively charged
pion-like state.

We consider ``hadron'' temperatures,
$T < T_{\rm qgp} \approx 160$~MeV, where the
quarks can be considered as a content of the hadrons, and assume that the
vast majority of quark configurations reflect the pion structure,
$\pi^+ = u\bar{d}$ and $\pi^- = d\bar{u}$.
With account for the isospin projections of the quarks, $+1/2$ for $u$
and $\bar d$ and $-1/2$ for $\bar u$ and $d$, and for the relations
$N_u = N_{\bar d} = N_\pi^{(+)}$, $N_{\bar u} = N_d = N_\pi^{(-)}$,
eq.~(\ref{eq:ni-pions-1}) can be rewritten as
\begin{equation}
N_I \,=\, \left(\frac 12 N_u + \frac 12 N_{\bar d}\right)
- \left(\frac 12 N_{\bar u} + \frac 12 N_d\right)
\,=\, N_u - N_d \,.
\label{eq:ni-pions-quarks}
\end{equation}
The free energy of the pion system with account for the quark content
reads
\begin{eqnarray}
F_\pi &=& - \, p_\pi V + \mu_u N_u + \mu_d N_d + \mu_{\bar u} N_{\bar u}
+ \mu_{\bar{d}} N_{\bar d}
\nonumber \\
&=& - \, p_\pi V \,+\, \mu_I^{(q)} \, \frac 12 \,
( N_u - N_d - N_{\bar u} + N_{\bar d} ) \,,
\label{eq:q-pion-free-energy-1}
\end{eqnarray}
where $\mu_I^{(q)}$ is the isospin chemical potential that characterizes the
quark configurations, i.e. $\mu_u = -\mu_{\bar u} = \mu_I^{(q)}/2$ and
$\mu_d = -\mu_{\bar d} = -\mu_I^{(q)}/2$, or
\begin{equation}
\mu_I^{(q)} \,=\, \mu_u - \mu_d \,.
\label{eq:q-chem-pot-11}
\end{equation}
With $N_u = N_{\bar d}$, $N_{\bar u} = N_d$ and
eq.~(\ref{eq:ni-pions-quarks}) the free energy takes the form
$F_\pi = - p_\pi V + \mu_I^{(q)} N_I$.
Comparing it with the free energy of the pion system, where only
$\pi^+$ and $\pi^-$ exist,
\begin{eqnarray}
F_\pi \,=\, - p_\pi V + \mu_\pi^{(+)} N_\pi^{(+)} + \mu_\pi^{(-)} N_\pi^{(-)} \,
=\, - p_\pi V + \mu_I^{(\pi)} N_I \,,
\label{eq:pion-free-energy}
\end{eqnarray}
we conclude that, if the quark configurations correspond to the pion
degrees of freedom, then
\begin{equation}
\mu_I^{(q)} \,=\, \mu_I^{(\pi)} \,=\, \mu_I \,.
\label{eq:isospin-pot-1}
\end{equation}
Since in an ideal gas the condensate is created when the chemical
potential equals the particle mass, the condensation of pions in terms of the
quark content begins at $\mu_I^{(q)} = \mu_I = m_\pi$.
This is also the convention of Son and Stephanov~\cite{son-2001}, who assign
to the light quarks chemical potentials of equal magnitude $|\mu_I|/2$ and
opposite sign, and find the onset of pion condensation at $|\mu_I| = m_\pi$
(see also \cite{anch-gnat-kondakova-2025} and references therein).

\medskip

In the lattice calculations
\cite{brandt-2021,brandt-2022,brandt-2018,brandt-2016,brandt-2017}
the isospin chemical potential of quarks is defined as
\begin{equation}
\mu_I \,=\, \frac 12 \, (\mu_u - \mu_d) \,,
\label{eq:iso-chem-pot-1}
\end{equation}
while the isospin number is $N_I = N_u - N_d$~\cite{brandt-2018} and the
baryon density is zero.
Then the free energy (\ref{eq:q-pion-free-energy-1}) takes the form
\begin{equation}
F_\pi \,=\, - Vp_\pi \,+\, 2\mu_I N_I \,,
\label{eq:euler-4}
\end{equation}
i.e. the pion chemical potential entering the Bose distribution is
$2\mu_I$, and the condition for the condensate creation in an ideal gas
becomes
\begin{equation}
\mu_I \,=\, \frac 12 m_\pi\,.
\label{eq:cond-cond-1}
\end{equation}
The difference between the two thresholds, $\mu_I = m_\pi$ and
$\mu_I = m_\pi/2$, is thus a normalization of the isospin chemical potential.

\section{Description of the lattice data}
\label{sec:fit-latt-data}

To describe the results of the lattice calculations
\cite{brandt-2016,brandt-2017,brandt-2018,brandt-2021,brandt-2022}, we use the mean-field
approach presented in section \ref{sec:thermal-cond-phase}.
We consider the pion-like system within two models, $\varphi^4$ and $\varphi^6$,
with a repulsive self-interaction.
We stress that the quasiparticles of the model are not the physical pions:
at the condensation threshold their effective mass is $M = m_\pi/2$, see
eq.~(\ref{eq:eff-mass(T=120)}) below, and they couple directly to the isospin
chemical potential in the lattice normalization (\ref{eq:iso-chem-pot-1}).
The model should therefore be regarded as a phenomenological parametrization
of the dependence of the lattice thermodynamics on $\mu_I$.
We aim to show the competition between these two models in fitting the lattice
data.

\begin{figure*}[t]
\includegraphics[width=0.48\textwidth]{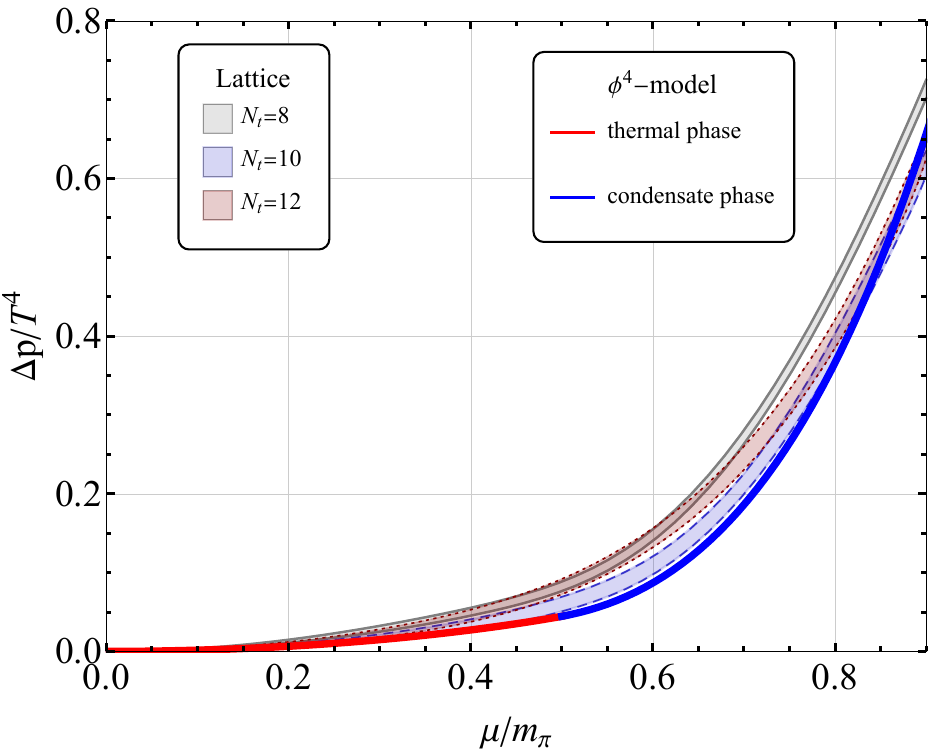}
\includegraphics[width=0.48\textwidth]{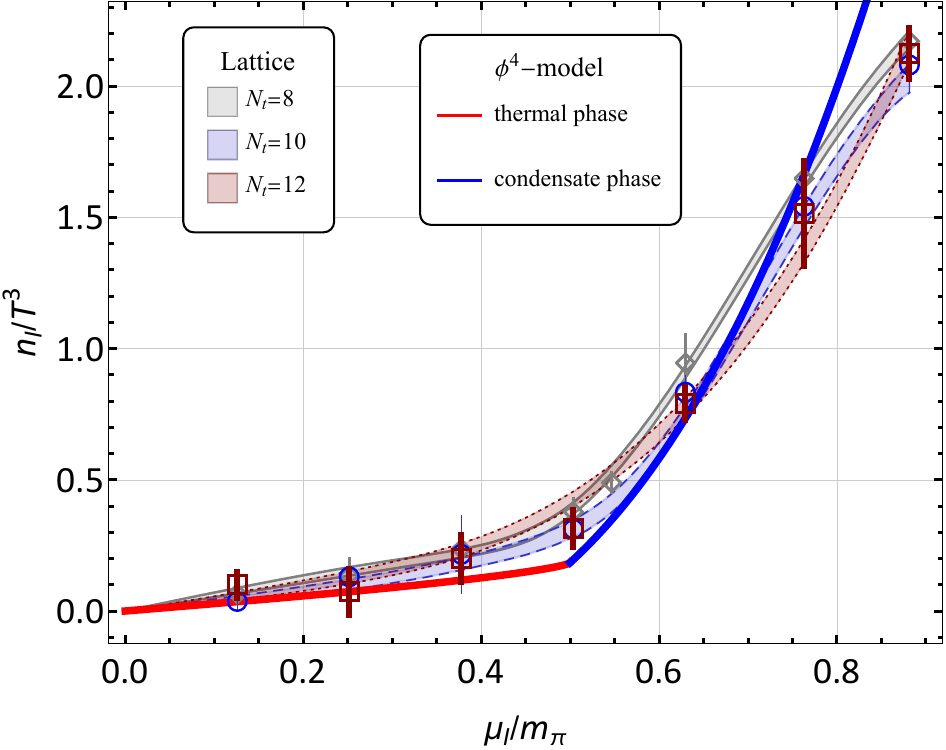}
\includegraphics[width=0.48\textwidth]{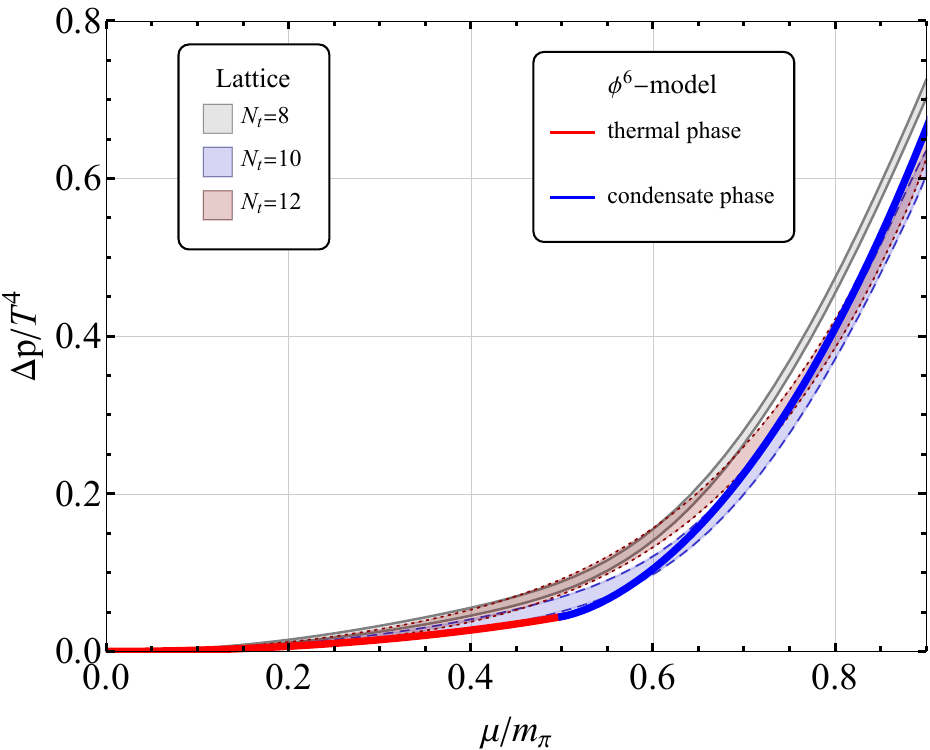}
\includegraphics[width=0.48\textwidth]{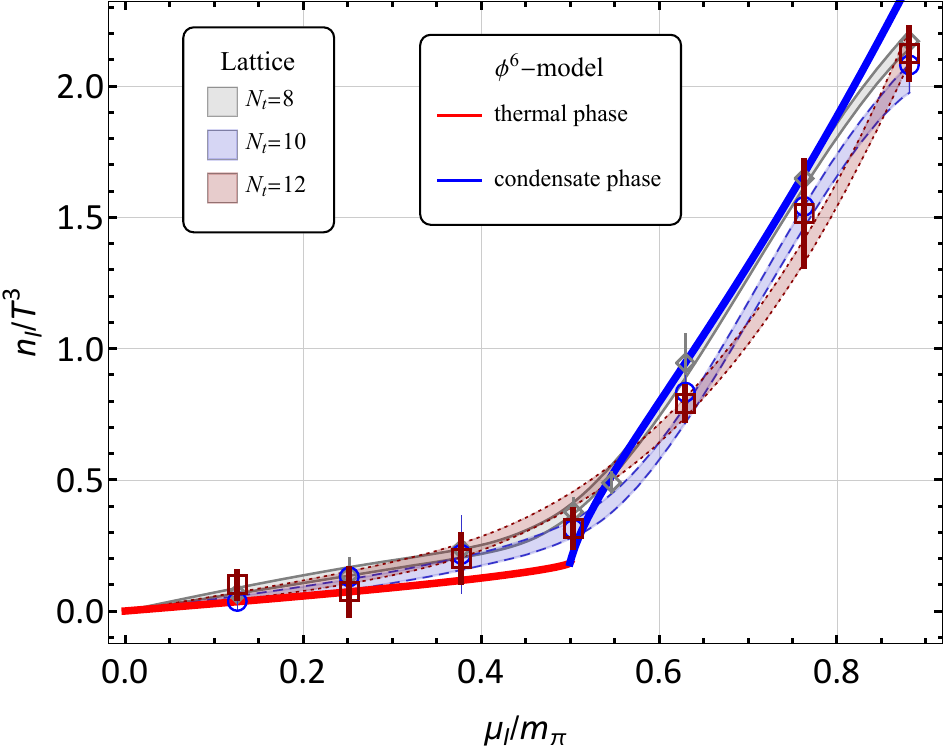}
\caption{\small Fit of lattice calculations results
\cite{brandt-2022} at $T = 120~$MeV for the quantity $\Delta p$ in
the $\varphi^4$ model (top left panel) and the $\varphi^6$ model
(bottom left panel). Description of lattice results
\cite{brandt-2022} at $T = 122~$MeV for $n_I$ in the $\varphi^4$
model (top right panel) and the $\varphi^6$ model (bottom right
panel). The red segment of the approximation curve refers to the
thermal phase, and the blue segment to the condensed phase.
 }
\label{fig:P-T122}
\end{figure*}

To fit the lattice data we use the same definitions that were adopted in
Refs.~\cite{brandt-2018,brandt-2021,brandt-2022}:
the pion mass value $m_\pi = 135$~MeV, $\mu_I = \frac 12 (\mu_u - \mu_d)$
and $N_I = N_u - N_d$.
That is, for the comparison below we use the lattice convention,
eq.~(\ref{eq:iso-chem-pot-1}).
We fit the data at the fixed temperature $T = 120$~MeV.
To be adequate to the lattice calculations
\cite{brandt-2021,brandt-2022,brandt-2018,brandt-2016,brandt-2017}
we adopt that at this temperature the phase transition to the condensed
phase occurs when $\mu_I = m_\pi/2$.
In accordance with our model, this means that at this temperature the effective
(thermal) mass $M$ of pions is
\begin{equation}
M(\sigma_{\rm cr}) \,=\, \frac 12 m_\pi \,,
\label{eq:eff-mass(T=120)}
\end{equation}
where the value of the scalar density $\sigma_{\rm cr}(T)$ belongs to the critical
curve.
The value $\sigma_{\rm cr}$ can be found from self-consistent equation
(\ref{eq:total-scalar-density-cond-2}) when $n_{\rm cond} = 0$:
\begin{equation}
\sigma_{\rm cr}(M,T) \,=\, \int \frac{d^3k}{(2\pi)^3\, 2\omega_k}
\left[ f_{\rm BE}\big(\omega_k,M\big) + f_{\rm BE}\big(\omega_k,-M\big)\right] \,,
\label{eq:det-sigma-cr}
\end{equation}
where $\omega_k = \sqrt{m_\pi^{2}/4 + \boldsymbol{k}^2}$ and we use condition of the
condensate creation of the positively charged particles, $\mu_I = M$, which is
valid in the condensed phase, starting from the critical point.
Here, in correspondence to Refs.~\cite{brandt-2021,brandt-2022},
in eq.~(\ref{eq:det-sigma-cr}) we use the value of the effective mass
$M = m_\pi/2$ at $T = 120$~MeV.
With these two values eq.~(\ref{eq:det-sigma-cr}) transforms into
a formula for the numerical calculation of the critical scalar density
$\sigma_{\rm cr}$.

As was declared in Section \ref{sec:complex-scalar-lattice} we fit the lattice
data using two approaches which correspond to two interaction Lagrangians
presented in eq.~(\ref{eq:lagrangian-lattice2}).
In accordance with the formalism given in Section \ref{sec:mf-approximation}
every type of interaction Lagrangian  determines
its own mean field (see eq.~(\ref{eq:definitions})):
$U(\sigma) = - \partial \mathcal{L}_{\rm int}(\sigma)/\partial \sigma$.
And further the effective mass $M$ of quasiparticles is determined as:
$M^2 = m_*^2 + U(\sigma)$, where $m_*$ is the bare mass
and $M(\sigma_{\rm cr}) = m_\pi/2$.
Hence for every approach at the critical point we get:
\begin{eqnarray}
\label{eq:m-phi4}
M^2 &=& m_*^2 + \lambda \sigma_{\rm cr} \qquad {\rm for} \quad \varphi^4~{\rm model} \,,
\\
M^2 &=& m_*^2 + b \sigma_{\rm cr}^2 \qquad {\rm for} \quad \varphi^6~{\rm model} \,.
\label{eq:m-phi6}
\end{eqnarray}
These relations give the connection between the value of the bare mass $m_*$ and
the interaction constant $\lambda$ or $b$, respectively.
To find interaction constant, we use eq.~(\ref{eq:pres-2}) to fit the lattice
pressure \cite{brandt-2022} as $\Delta p(T,\mu_I) = p(T,\mu_I) - p(T,0)$.
For the best fit of the $\Delta p$ we find: $\lambda \approx 0.4934$,
$b m_\pi^2 \approx 0.5673$.
In accordance with eqs.~(\ref{eq:m-phi4}) and (\ref{eq:m-phi6}) the corresponding
fitted bare masses are $m_*=m_\pi/2.22$ for the $\varphi^4$ model and
$m_*=m_\pi/2.021$ for the $\varphi^6$ model.
This fit of the lattice pressure is shown in Fig.~\ref{fig:P-T122}, in the left
panels.
To our understanding, the result, that the bare mass of particles in both
approaches is approximately two times smaller
than the pion mass $m_\pi = 135$~MeV, is a consequence of the normalization
of the chemical potential $\mu_I$ for quarks adopted in
\cite{brandt-2021,brandt-2022,brandt-2018,brandt-2016,brandt-2017},
see eq.~(\ref{eq:iso-chem-pot-1}).

\begin{figure*}[t]
\includegraphics[width=0.48\textwidth]{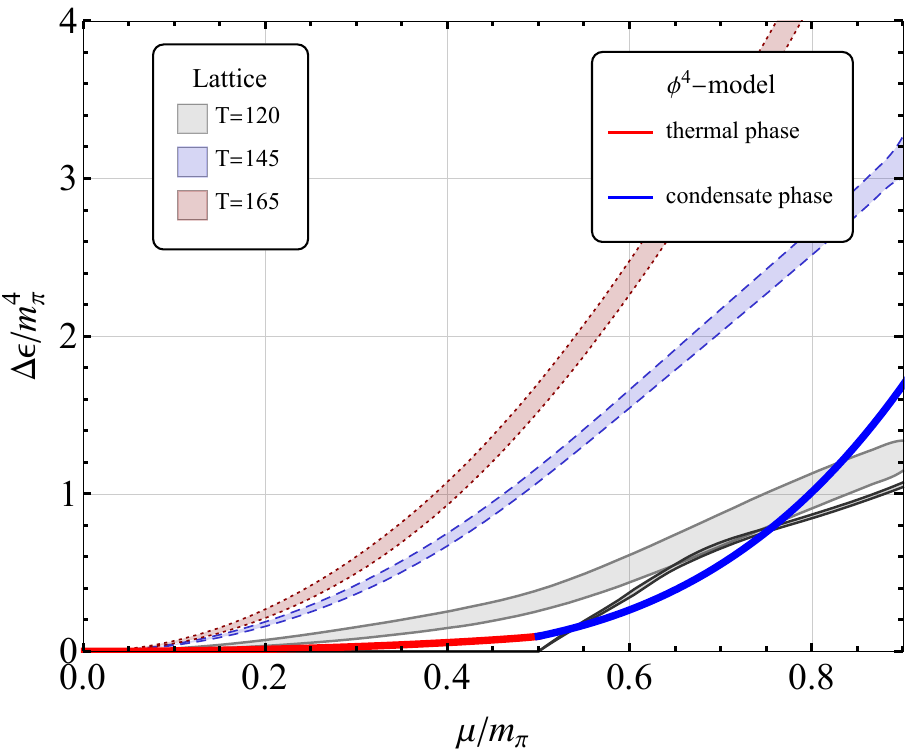}
\includegraphics[width=0.48\textwidth]{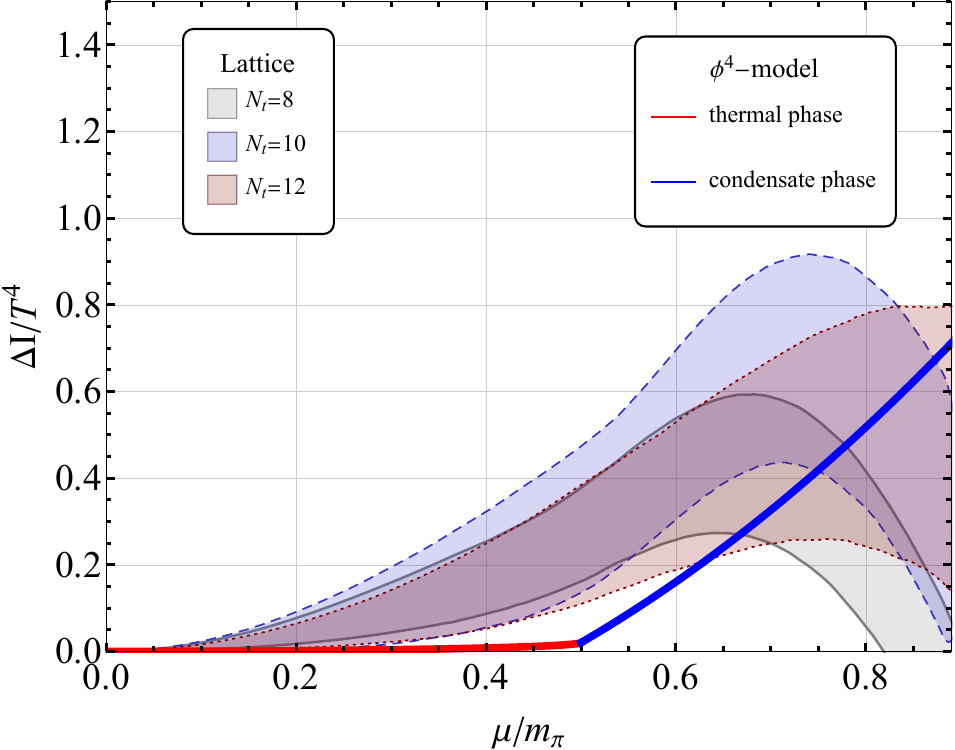}
\includegraphics[width=0.48\textwidth]{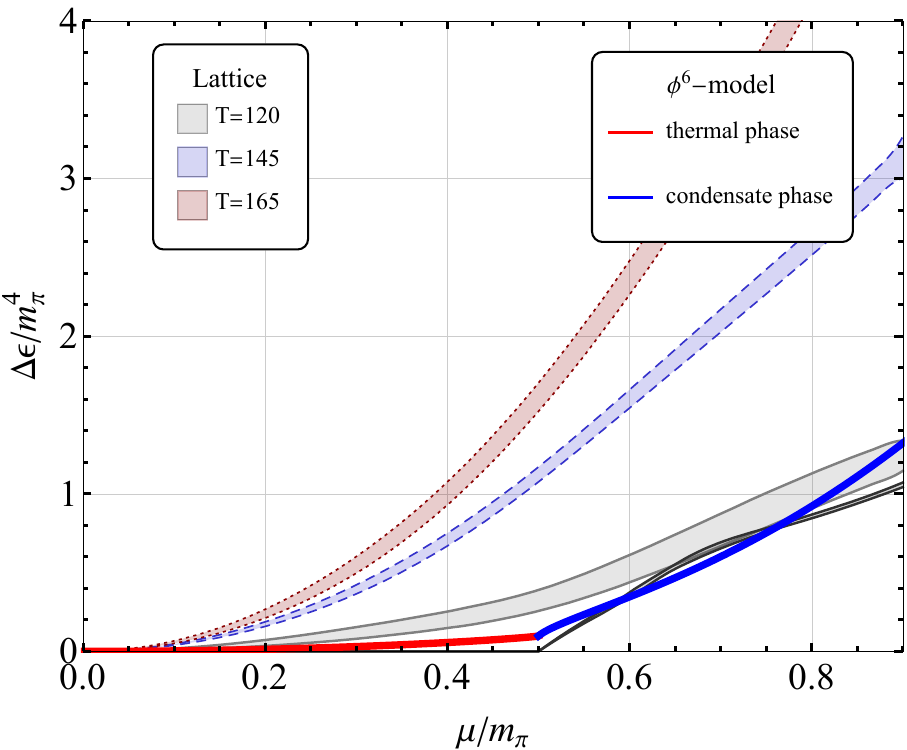}
\includegraphics[width=0.48\textwidth]{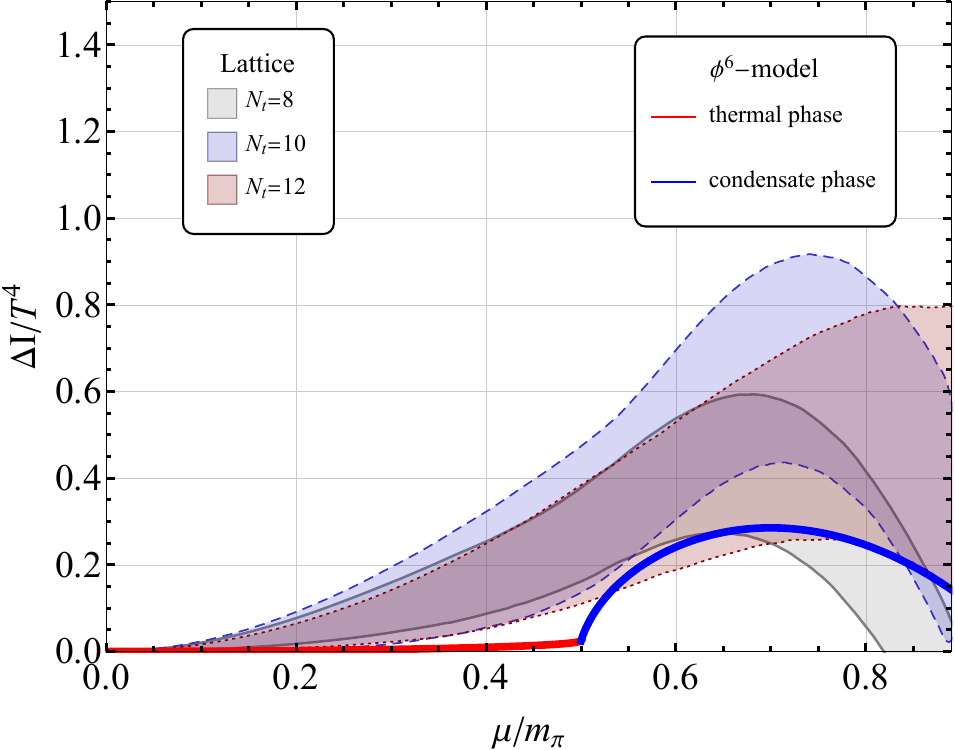}
\caption{\small Description of lattice calculation results
\cite{brandt-2021,brandt-2022} at $T = 120~$MeV
for the quantities $\Delta \varepsilon$ and $\Delta I$ in the $\varphi^4$
model (top panels) and the $\varphi^6$ model (bottom panels).
The red segment of the fit curve corresponds to the thermal phase, and
the blue segment to the condensed phase.
 }
\label{fig:delta-I-T122}
\end{figure*}

In the thermal region, the chemical potential is obtained by solving the
thermal self-consistency equations at fixed $T$ and $n_I$.
In the condensed region it is constrained by $\mu_I = M(T, n_I)$, with the
effective mass determined from the coupled condensate equations.
For this reason, at a fixed temperature $T = 120$~MeV, we parameterize
all quantities in terms of the isospin density $n_I$:
$n_I < n_{\rm Icr}$ for the thermal phase and $n_I \ge n_{\rm Icr}$ for the
condensed phase, where $n_{\rm Icr}$ is obtained by performing the integration
in eq.~(\ref{eq:charge-density-3}) under the conditions $n_{\rm cond} = 0$
and $M = m_\pi/2$.
In all panels in Figs.~\ref{fig:P-T122} and \ref{fig:delta-I-T122}
the $X$-axis is a chemical potential, which is taken as
solution of eqs.~(\ref{eq:isospin-density}), (\ref{eq:total-scalar-density})
for fixed temperature in the thermal phase;
and in the condensed phase a chemical potential equals to the effective mass,
$\mu_I = M(T, n_I)$, which is taken as solution of
eqs.~(\ref{eq:charge-density-3}), (\ref{eq:total-scalar-density-cond-2}).
In this sense the calculations are carried out in the Extended Canonical
Ensemble.

The calculation of all thermodynamic quantities was carried out using the
interaction constants $\lambda \approx 0.4934$ and $b m_\pi^2 \approx 0.5673$
obtained by fitting the pressure.
The results of the calculation of the isospin density $n_I$
(see, eqs.~(\ref{eq:isospin-density}), (\ref{eq:charge-density-3}))
at $T = 122$~MeV together with the lattice data \cite{brandt-2022} are shown
in Fig.~\ref{fig:P-T122}, in the right panels.
One can see a satisfactory description of the lattice calculation results,
where the $\varphi^6$ being more accurate.
The results of the calculation at $T = 120$~MeV of the energy density
in the form
\begin{equation}
\Delta \varepsilon(T,\mu_I) \,=\, \varepsilon(T,\mu_I) - \varepsilon(T,0) \,,
\label{eq:delta-epsilon}
\end{equation}
and the trace-anomaly
\begin{equation}
I \,=\, \varepsilon - 3p \,=\, I(T,0) + \Delta I(T,\mu_I) \,,
\label{eq:trace-anom}
\end{equation}
are shown in Fig.~\ref{fig:delta-I-T122} in left and right panels together with
the lattice data \cite{brandt-2022}.
In all figures the red segment of the curve relates to the thermal phase and
the blue segment of the curve relates to the condensed phase.
We see quite good description of the lattice data.
However, comparison of two model results leads to conclusion, that the
$\varphi^6$ model describes better the results of the lattice calculations.

The agreement is nontrivial because the quantities shown in the different
panels are not independently fitted: once the interaction parameter $\lambda$
or $b$ is fixed from the pressure, the isospin density, energy density, and
trace anomaly follow from the same thermodynamically consistent framework.

The comparison therefore provides a test of both the interaction model and
the thermodynamic formulation.
In particular, the thermal and condensed branches match continuously
at the transition.

\begin{figure*}[t]
\includegraphics[width=0.48\textwidth]{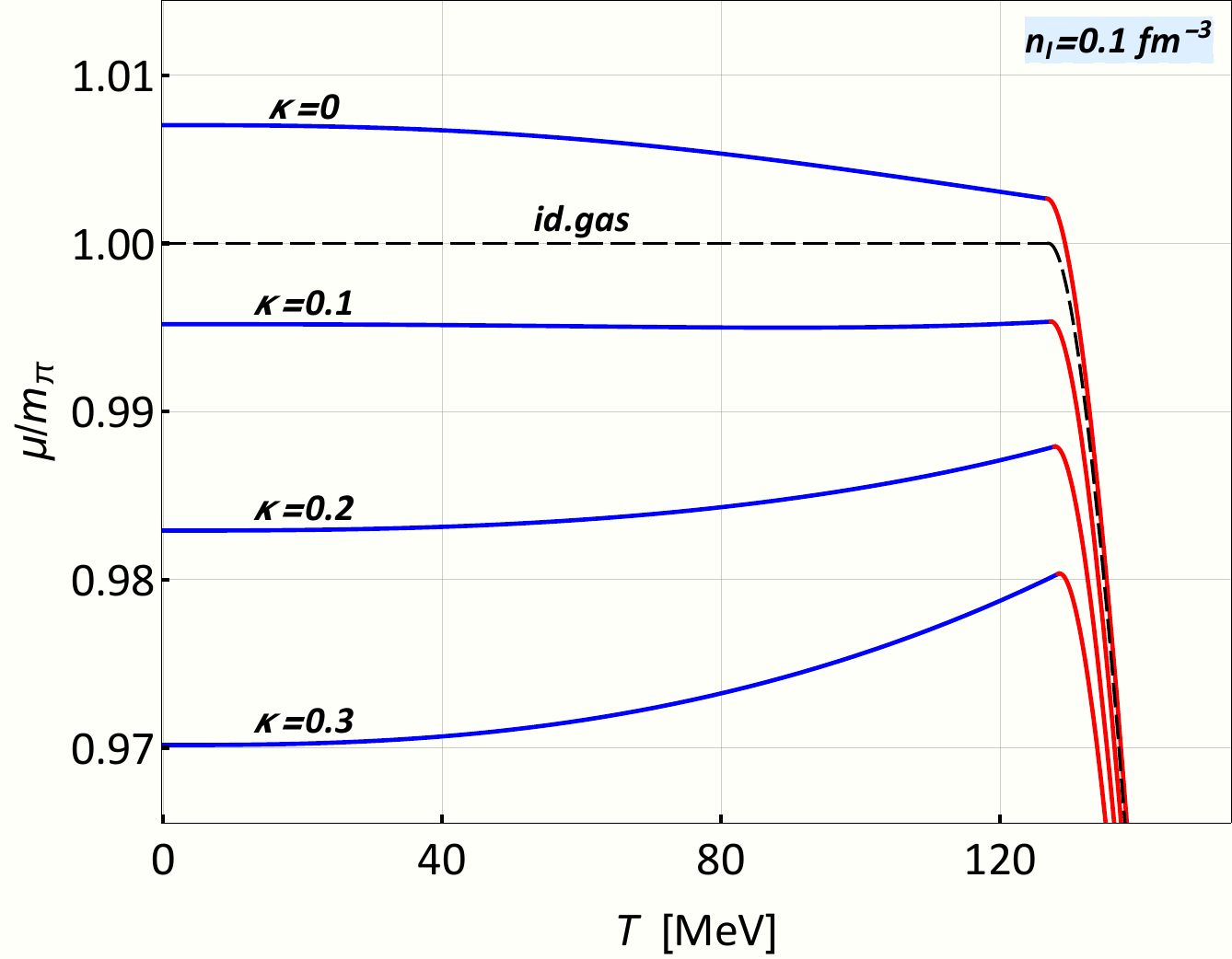}
\caption{\small
Dependence of the $\mu_I$ on the temperature at $n_I = 0.1$~fm$^{-3}$
for different intensities of the attraction between particles that is
parameterized by the dimensionless parameter $\kappa$, the coefficient
that determines the strength of the attraction (see Ref.~\cite{mishustin-anchishkin-2019}):
a. $\kappa = 0$ -- no attraction, there is only repulsion between particles
(top curve);
b. an ideal gas -- the black dashed curve;
c. an increasing intensity of attraction corresponds to the curves labeled
by an increasing value of the $\kappa = 0.1,\, 0.2,\, 0.3$.
The blue segment of the curves corresponds to the condensed phase,
the red segment to the thermal phase.
}
\label{fig:chem-potential}
\end{figure*}

It is necessary to point out, that for a boson system with repulsion only, when
$M \ge m_* \quad \rightarrow \quad \mu_I \ge m_*$ in the condensed phase,
the thermodynamic quantities obtained in this way can be presented as
functions of $\mu_I$,
since at fixed temperature the correspondence between $n_I$ and $\mu_I$
is one-to-one.
In this case it is necessary to account for a transformation of the dispersion
law in the condensed phase:
\begin{equation}
\omega_k = \sqrt{M^2 + \bs k^2}
\quad \rightarrow \quad
\omega_k = \sqrt{\mu_I^2 + \bs k^2} \,.
\label{eq:dispersion-mu}
\end{equation}
And for the same reason the set of eqs.~(\ref{eq:charge-density-3}),
(\ref{eq:total-scalar-density-cond-2}) must be rewritten as
\begin{eqnarray}
\label{eq:charge-density-3mu}
&& n_I = n_{\rm cond}(T,n_I)
+ \int \frac{d^3k}{(2\pi)^3}\, \left[f_{\rm BE}\big(\omega_k, \mu_I\big)
- f_{\rm BE}\big(\omega_k,- \mu_I\big) \right] \,,
\\
&& \sigma =  \frac{\dis n_{\rm cond}}{\dis 2M}
+ \int \frac{d^3k}{(2\pi)^3\, 2\omega_k}\, \left[f_{\rm BE}\big(\omega_k, \mu_I\big)
+ f_{\rm BE}\big(\omega_k,- \mu_I \big) \right] \,,
\label{eq:total-scalar-density-cond-2mu}
\end{eqnarray}
where, $\sigma = (\mu_I^2 - m^2)/\lambda$ for $\varphi^4$ model and
$\sigma = \sqrt{(\mu_I^2 - m^2)/b}$ for $\varphi^6$ model.
It is seen that such a transformation of the dynamical variable,
$\sigma \rightarrow \mu_I$, is possible only when $\mu_I \ge m$, i.e.
it is allowed only for the repulsive interaction between particles, $M \ge m$.

In the boson system in the case of attraction between particles it is impossible
to parameterize the thermodynamic quantities by the chemical potential,
since the correspondence between $\mu_I$ and the thermodynamic states becomes
two-valued: two different thermodynamic states correspond to one value of $\mu_I$.
This was schematically shown in Fig.~\ref{fig:chem-potential}.
It is seen that for the top curve, i.e. there is  only repulsion, all states
can be characterized unequivocally by the pair $(T,\mu_I)$.
It is not the case for the ideal gas, where in the condensed phase all
states in the interval $0 \le T \le T_{\rm c}$ have the same chemical potential
$\mu_I = m_\pi$.
When an attraction is ``switched on'', for all curves below the ideal gas
one definite value of the chemical potential corresponds to a couple of the
thermodynamic states -- one state in the condensed phase (blue segment) and
another in the thermal phase (red segment), see Fig.~\ref{fig:chem-potential}.

\begin{figure*}[t]
\includegraphics[width=0.3\textwidth]{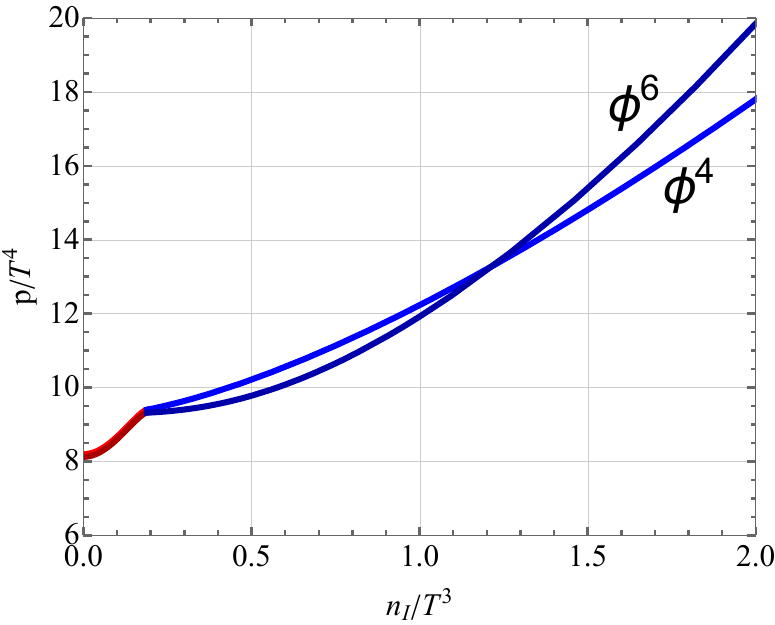}
\includegraphics[width=0.3\textwidth]{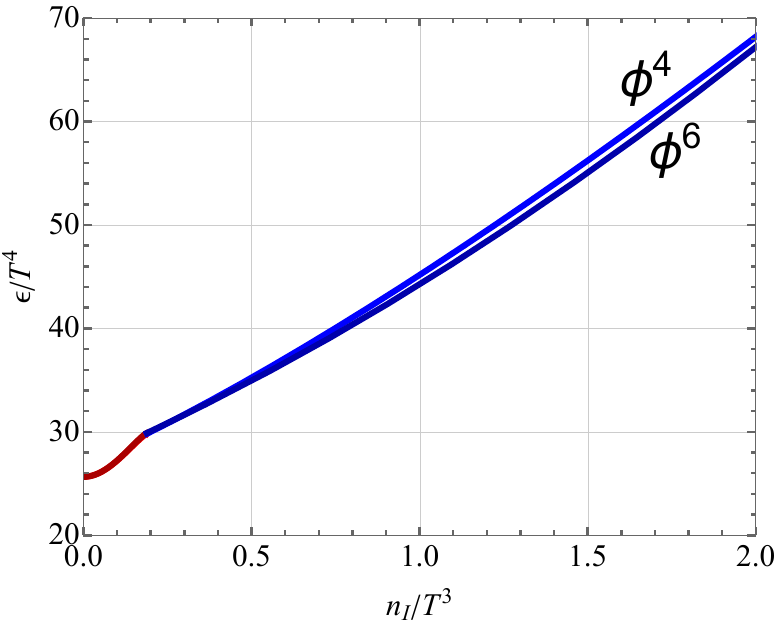}
\includegraphics[width=0.3\textwidth]{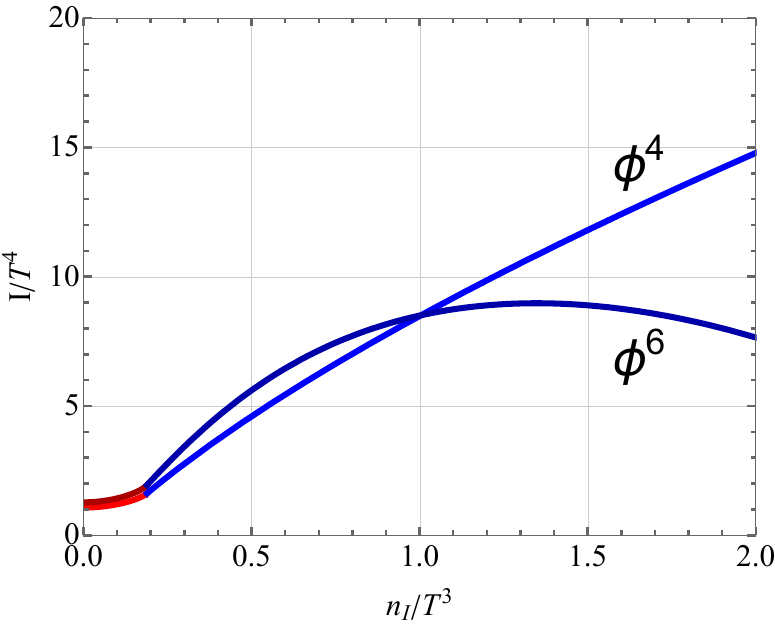}
\caption{\small
Dependence of the pressure (left panel), energy density (central panel),
and trace anomaly (right panel) at $T = 120~$MeV on the isospin density $n_I$
calculated for $\varphi^4$ and $\varphi^6$ models.
The red segment of the each curve corresponds to the thermal phase, and
the blue segment to the condensed phase.
 }
\label{fig:p-eps-I-T120}
\end{figure*}

For parametrization of the pressure, energy density, and trace anomaly, for the
same interaction constant and in the same framework as it was done before,
one can use the real physical quantity -- isospin density $n_I$.
The results of the calculation for $\varphi^4$ and $\varphi^6$ models are shown in
Fig.~\ref{fig:p-eps-I-T120}.
In this figure the beginning of dependencies on every panel starts from the
following values:
\[ p(T,\mu_I = 0) = p(T,n_I = 0) \approx 8.2\, T^4 \,, \]
\[ \varepsilon(T,\mu_I = 0) = \varepsilon(T,n_I = 0) \approx 25.8\, T^4 \,, \]
\[ I(T,\mu_I = 0) = I(T,n_I = 0) \approx 1.2\, T^4 \,. \]
These values were set to zero for the dependencies of the same
thermodynamic quantities on $\mu_I$
in Figs.~\ref{fig:P-T122}, \ref{fig:delta-I-T122}.

\section{Conclusions}
\label{sec:conclusion}

We have studied an interacting relativistic system of charged bosons at finite
temperature and fixed isospin density within a thermodynamically consistent
mean-field approach.
Repulsive $\varphi^4$ and $\varphi^6$ self-interactions generate a density-dependent
quasiparticle mass and an excess-pressure contribution.
The thermodynamic properties are formulated in the Extended Canonical Ensemble,
in which the isospin density is the independent conserved variable.

A characteristic feature of the condensed phase is that the isospin chemical
potential is constrained by the Bose-condensation condition, $\mu_I = M(\sigma)$,
where the effective mass is determined self-consistently by the scalar density.
For purely repulsive interactions the results can equally be presented
as functions of $\mu_I$, since at fixed temperature the correspondence between
$n_I$ and $\mu_I$ is one-to-one.
The ECE formulation provides a convenient description of the system across the
thermal-to-condensed transition and becomes necessary for an ideal gas or in
the presence of attraction, where the value of $\mu_I$ does not determine the
thermodynamic state uniquely.

We have also analyzed the relation between the isospin chemical potentials
defined for pion and quark degrees of freedom.
The numerical value of the chemical potential at the condensation threshold
depends on its normalization.
In the convention $\mu_I = \mu_u - \mu_d$ the ideal-gas condition is $\mu_I = m_\pi$,
whereas in the convention used in lattice-QCD calculations
\cite{brandt-2021,brandt-2022,brandt-2018,brandt-2016,brandt-2017},
$\mu_I = (\mu_u - \mu_d)/2$, it becomes $\mu_I = m_\pi/2$.
Making this distinction explicit is essential for a consistent comparison
between the pion and quark descriptions.

Finally, we applied the mean-field equations, regarded as a
phenomenological parametrization, to lattice-QCD results
at $T = 120$~MeV.
Both the $\varphi^4$ and $\varphi^6$ models reproduce the main behavior of the
pressure, isospin density, energy density, and trace anomaly.
In each model only one parameter, the interaction constant, is fitted to
the lattice pressure; the isospin density, energy density, and trace anomaly
then follow without additional fitting.
Within the present parametrization, the $\varphi^6$ model provides the better
overall description of the lattice results \cite{brandt-2022}.
At the same time, the success of the presented phenomenological interacting pion-like
models suggests that, at ``hadron'' temperatures, at least at $T = 120$~MeV,
and at the considered isospin densities, the vast majority of quark
configurations reflect pion structure,
with $n_{\bar d} = n_u$ and $n_{\bar u} = n_d$.

These results support the use of a thermodynamically consistent
interacting-boson model as an effective description of finite-isospin
hadronic matter across the Bose-condensation transition.

\section*{Acknowledgements}
The work is supported by the National Research Foundation of Ukraine
under the grant number 2025.07/0461.
Authors thank A.~Korchin for useful and fruitful discussions.

\end{document}